\documentclass[reprint,aps,prb.superscriptaddress,amsmath,amssymb,twocolumn,showpacs]{revtex4-1}

\usepackage{graphicx} 
\usepackage{dcolumn} 
\usepackage{bm} 
\usepackage{amsmath}
\usepackage{braket}
\usepackage{natbib}
\usepackage{color}%
\usepackage{ulem}
\usepackage{lineno}

\newcommand*\patchAmsMathEnvironmentForLineno[1]{%
  \expandafter\let\csname old#1\expandafter\endcsname\csname #1\endcsname
  \expandafter\let\csname oldend#1\expandafter\endcsname\csname end#1\endcsname
  \renewenvironment{#1}%
     {\linenomath\csname old#1\endcsname}%
     {\csname oldend#1\endcsname\endlinenomath}}%
\newcommand*\patchBothAmsMathEnvironmentsForLineno[1]{%
  \patchAmsMathEnvironmentForLineno{#1}%
  \patchAmsMathEnvironmentForLineno{#1*}}%
\AtBeginDocument{%
\patchBothAmsMathEnvironmentsForLineno{equation}%
\patchBothAmsMathEnvironmentsForLineno{align}%
\patchBothAmsMathEnvironmentsForLineno{flalign}%
\patchBothAmsMathEnvironmentsForLineno{alignat}%
\patchBothAmsMathEnvironmentsForLineno{gather}%
\patchBothAmsMathEnvironmentsForLineno{multline}%
}

\begin{document}


\title{Phonon rotoelectric effect}




\author{Masato Hamada}
\author{Shuichi Murakami}
\affiliation{Department of Physics, Tokyo Institute of Technology, 2-12-1 Ookayama, Meguro-ku, Tokyo 152-8551, Japan}



\date{\today}

\begin{abstract}
In crystals with time-reversal symmetry but without inversion symmetry, 
the phonon angular momentum can be generated by the temperature gradient, and it is called phonon thermal Edelstein effect. 
On the other hand, when both symmetries are broken and their product is conserved, the phonon angular momentum 
for any phonon modes at any wave vectors
vanishes, and the phonon thermal Edelstein effect does not occur. 
In this paper, we propose another mechanism of generation of the phonon angular momentum. 
We show that in such crystals the electric field generates the phonon angular momentum, via the lattice distortion due to the electric field.
This effect is in the same symmetry class with the magnetoelectric effect, and we call this effect phonon rotoelectric effect.
We discuss the temperature dependence of the phonon angular momentum generated by the temperature gradient and by the electric field in the high- and the low-temperature limits.
\end{abstract}



\maketitle



\section{Introduction}


In recent years, interesting properties of phonons have been reported such as phonon Hall effect
\cite{PHEexp01, PHEexp02, PHEexp03, PhysRevLett.100.145902, PhysRevLett.105.225901} 
and topological nature of phonon systems
\cite{SC01,SC02,mechanicalgraphene01,mechanicalgraphene02,mechanicalgraphene03,PhysRevB.96.064106}. 
In the phonon Hall effect, phonons couple with a magnetic field via the spin-phonon interaction, and then a flow of phonons having microscopic angular momentum can be generated
\cite{PAM01}. 
The phonon angular momentum represents microscopic local rotations around the equilibrium positions of atoms in the crystal lattices. 
Various phenomena related with the phonon angular momentum have been proposed,
such as correction to the Einstein-de Haas effect
\cite{PAM01}, 
spin relaxation
\cite{PAM02, PhysRevB.97.174403, PhysRevLett.121.027202}, 
orbital magnetization of phonons
\cite{PhysRevMaterials.3.064405}, 
and conversion between magnons and phonons
\cite{NatPhys.14.500, PhysRevB.92.214437}. 
In systems without inversion symmetry, phonon modes have chirality
\cite{PAM03}. 
In particular, in the valleys of the phonon band structure in momentum space, the angular momenta of the chiral phonons are quantized and their excitation by intervalley scattering of electrons using circular polarized light has been proposed, and the chiral phonons has been experimentally observed in monolayer tungsten diselenide
\cite{chiralphononexp01}. 
Moreover, in mechanical graphene and sonic crystals, topological edge modes of phonons have been reported 
based on the chirality of phonons
\cite{SC01,SC02,PhysRevB.96.064106}.
On the other hand, methods of generation of the phonon angular momentum have been proposed by several means, such as an external magnetic field~\cite{PAM01, PhysRevMaterials.1.014401}, circular polarized light~\cite{PAM03, chiralphononexp01}, infrared excitation~\cite{PhysRevMaterials.3.064405, nova2017}, a rigid-body rotation of crystals~\cite{Wang_2015, PhysRevB.96.064106}, and a temperature gradient~\cite{Masato02}.
The classification of materials with phonon angular momentum for time-reversal symmetry, inversion symmetry, and their product is reported~\cite{coh2019classification}.

The distribution of the phonon angular momentum in momentum space is constrained by
crystallographic symmetries and time-reversal symmetry. 
In particular, the phonon angular momentum vanishes for any phonon modes when both time-reversal and inversion symmetries are present.
In the previous study
\cite{PAM01}, 
the phonon angular momentum 
is shown to be nonzero in systems without time-reversal symmetry.
In our previous work, we reported that the phonon angular momentum is generated by temperature gradient in crystals without inversion symmetry, such as wurtzite gallium nitride, tellurium and selenium, and we call this effect phonon thermal Edelstein effect
\cite{Masato02}. 
On the other hand, in systems with the product of time-reversal and inversion symmetries, the phonon angular momentum of every phonon mode becomes zero at any wave vectors. 
Hence, the total phonon angular momentum in equilibrium is zero. 
Moreover, the phonon angular momentum due to phonon thermal Edelstein effect cannot be generated.
In this paper, we theoretically propose generation of the phonon angular momentum by an electric field in systems with neither time-reversal nor inversion symmetries. 
We first discuss this effect from the view point of 
magnetic point group symmetries.
Then, we show generation of the phonon angular momentum by an electric field using a two-dimensional spring-mass model with spin-phonon interaction, and discuss that it is due to the lattice distortion by the electric field.
Moreover, we discuss differences of two phenomena of generation of phonon angular momentum, one by the temperature gradient and the other by the electric field.

\section{Phonon angular momentum and symmetry}
The angular momentum of nuclei motions in crystals consists of two parts; a rigid-body rotation of the whole crystal and microscopic local rotations of the atoms around their equilibrium positions, and the latter is called phonon angular momentum
\cite{PAM01}. 
In equilibrium, the phonon angular momentum per unit volume is represented as 
\begin{gather}
J^{\rm ph}_{\alpha} = \frac{1}{V} \sum_{\bm{k},\sigma} l_{\sigma,\alpha}(\bm{k})\left(f_0(\omega_{\sigma}(\bm{k})) + \frac{1}{2}\right),~ \alpha=x,y,z, \label{pamj} \\
l_{\sigma,\alpha}(\bm{k}) = \hbar \epsilon_{\sigma}^{\dagger}(\bm{k})M_{\alpha}\epsilon_{\sigma}(\bm{k}), \label{paml}
\end{gather}
where $f_0(\omega_{\sigma}(\bm{k})) = 1/(e^{\hbar\omega_{\sigma}(\bm{k})/k_{\rm B}T}-1)$ is the Bose distribution function, $T$ is the temperature, and $V$ denotes the sample volume.  
$\epsilon_{\sigma}(\bm{k})$ is displacement polarization vector at the wave vector $\bm{k}$ with a mode index $\sigma$, and $\omega_{\sigma}(\bm{k})$ is the eigenfrequency of a phonon mode. 
These quantities $\epsilon_{\sigma}(\bm{k})$ and $\omega_{\sigma}(\bm{k})$ constitute a solution of the eigenmode equation for phonons $D(\bm{k})\epsilon_{\sigma}(\bm{k}) = \omega_{\sigma}^{2}(\bm{k})\epsilon_{\sigma}(\bm{k})$, where $D$ is the dynamical matrix. 
The matrix $M_{\alpha}$ is the tensor product of the $N\times N$ unit matrix for a unit cell with $N$ atoms and a generator of $SO(3)$ rotations,
and is given by $(M_{\alpha})_{\beta\gamma}=I_{N\times N}\otimes (-i)\varepsilon_{\alpha\beta\gamma}~(\alpha,\beta,\gamma=x,y,z)$. 
$\bm{l}_{\sigma}(\bm{k})$ in Eq.~(\ref{paml}) is the phonon angular momentum of a phonon mode $\sigma$ at wave vector $\bm{k}$. 
We note that a nonzero angular momentum is not necessarliy associated with a superposition of degenerated two phonon modes as studied in Refs.~\cite{PAM02, PhysRevMaterials.3.064405, nova2017}, but can appear in a single phonon mode by acquiring elliptic motions of nuclei.
The phonon angular momentum of each phonon mode satisfies the following relation; in the system with time-reversal (inversion) symmetry, it is an odd (even) function of the wave vector $\bm{k}$: $\bm{l}_{\sigma}(\bm{k}) = \mp \bm{l}_{\sigma}(-\bm{k})$. 
Thus, to make the phonon angular momentum nonzero, at least one of these symmetries must be broken.  
In the crystals without inversion symmetry, the phonon angular momentum sums up to zero in equilibrium, but it becomes nonzero by the temperature gradient
\cite{Masato02}.
Here, we focus on systems with neither of these symmetries, namely as magnetic crystals without inversion symmetry.
First we propose the phonon thermal Edelstein effect in magnetic insulators without the product of  time-reversal and inversion symmetries.
Next, we propose the new effect to generate the phonon angular momentum in magnetic insulator with the product of time-reversal and inversion symmetries.

\subsection{Phonon thermal Edelstein effect in magnetic insulators}
We focus on systems with neither of these symmetries, namely as magnetic crystals without inversion symmetry, and we discuss the extension of the phonon thermal Edelstein effect from the nonmagnetic insulators to the magnetic insulators. 
Note that in the phonon sysmtes without time-reversal symmetry, the eigenvalue equation for phonons couples between positive-frequency modes $(\sigma >0)$ with negative-frequency modes $(\sigma <0)$.
Nevertheless, they represent the same mode via $\omega_{\sigma}(\bm{k})=-\omega_{-\sigma}(-\bm{k})$, and therefore, the summation in Eq.~(\ref{pamj}) is limited to positive-frequency modes $(\sigma >0)$~\cite{PAM01}.
When the temperature gradient is applied to a system with neither of these symmetries, the phonon angular momentum $J_{\alpha}^{\rm ph}$ consists of two terms up to the linear order in the temperature gradient; an equilibrium term $J_{\alpha}^{\rm equil.}$ and a term proportional to the temperature gradient, and it is given by
\begin{equation}
J^{\rm ph}_{\alpha} = J_{\alpha}^{\rm equil.} + \alpha_{\alpha\beta} \frac{\partial T}{\partial x_{\beta}}, \label{pa}
\end{equation}
where $\alpha$ and $\beta$ are real-space coordinates, and $\alpha_{\alpha\beta}$ represents a response tensor to the temperature gradient.
Because the phonon angular momentum of each mode is no longer an even or odd function of the wave vector $\bm{k}$ when neither of these symmetries are present, the phonon angular momentum in equilibrium does not cancel between $\bm{k}$ and $-\bm{k}$. 
In addition, the phonon thermal Edelstein effect is also allowed. 
Note that the response tensor $\alpha$ for the phonon thermal Edelstein effect in magnetic crystals is determined by their magnetic point groups. 
The response tensor $\alpha$ 
is nonzero for 40 magnetic point groups, and a full list of the form of the axial tensor $\alpha_{\alpha\beta}$ is available from Ref.
\cite{1962series}.

\subsection{Phonon rotoelectric effect in magnetic insulators}
On the other hand, when both inversion and time-reversal symmetries are broken and their product is conserved, the phonon angular momentum for each mode $\bm{l}_{\sigma}(\bm{k})$ becomes zero at any wave vector. 
Therefore, the total phonon angular momentum becomes zero. 
In this case, the phonon thermal Edelstein effect does not occur because the change of the phonon distribution does not contribute to the phonon angular momentum, owing to $\bm{l}_{\sigma}(\bm{k})=0$. 
Here, we theoretically propose another mechanism of generation of the phonon angular momentum. 
A similar mechanism is known in multiferroic materials, called magnetoelectric effect
\cite{Cr2O3theory1}. 
The magnetoelectric effect
is expressed as $M_{\alpha} = \alpha_{\alpha \beta}^{\rm ME} E_{\beta}, P_{\alpha}= \alpha_{\beta\alpha}^{\rm ME} B_{\beta}$, where $M,P,B,E$ are a magnetization, a polarization, a magnetic field, and an electric field, respectively.
This cross-correlations are allowed by absence of both time-reversal and inversion symmetries. 
Because the magnetization and the phonon angular momentum share the same symmetry properties,
we show that the phonon angular momentum is generated by an electric field, in analogy to the magnetoelectric effect. 

In order to couple the magnetization with phonons, we introduce the spin-phonon interaction
\cite{SPI01,SPI02,SPI03,SPI04,SPI05}, 
which couples localized spins and the phonon angular momentum, and this is represented as 
\begin{align}
H_{\rm{SPI}} = -g\sum_{l\kappa} \bm{S}_{\kappa}\cdot (\bm{u}_{l\kappa}\times m_{\kappa}\dot{\bm{u}}_{l\kappa}), \label{spi}
\end{align}
where $g$ is a coupling constant, $\bm{S}_{\kappa}$ is the magnetization of localized spins at the $\kappa$th atom in the unit cell, $m_{\kappa}$ is the mass of the $\kappa$th atom in the unit cell, and $\bm{u}_{l\kappa}$ is the displacement vector of the $\kappa$th atom in the $l$th unit cell. 
This interaction is known as Raman spin-phonon interaction.
The physical mechanism is the coupling of the charged ions to the magnetic field created by the localized spins.
Because this interaction Eq.~(\ref{spi}) works like a Lorentz force to the nulclei, the motion of the nuclei is deflected by the Lorentz force and the phonon angular momentum is generated.
We note that the role of the spin-phonon interaction is not to split left- and right-circularly polarized phonons. Phonon modes at a general $\bm{k}$ points are nondegenerate, and when the spin-phonon interaction is taken into account, the motions of the nuclei become elliptic, and the phonon modes acquire angular momenta. 

Thus in an analogy with the magnetoelectric effect, 
we expect that an electric field generates the phonon angular momentum. 
It is represented as
\begin{align}
J_{\alpha}^{\rm{ph}} = \beta_{\alpha\beta}E_{\beta}, \label{pmee}
\end{align}
where $\bm{E}$ is an electric field and $\beta_{\alpha\beta}$ is a response tensor, and we call this effect phonon rotoelectric effect.
The response tensor $\beta_{\alpha\beta}$ is an axial tensor, and it is nonzero for $18$ magnetic point groups with the product of time-reversal and inversion operations and $40$ magnetic point groups without it.

We note that the phonon rotoelectric effect in this paper is the coupling between the microsocpic local rotation of atoms and the electric field.
It is different from the rotoelectricity~\cite{rotoelectricity}, which means generation of the electric polarization  by the static rotation angle of octahedra in materials forming the perovskite structure.
Thus, we call the effect proposed in this paper ``phonon rotoelectric effect'' to emphasize that it induces phonons with angular momentum.

\section{spring-mass model in two dimensions}
\begin{figure}
\includegraphics[width=8.5cm]{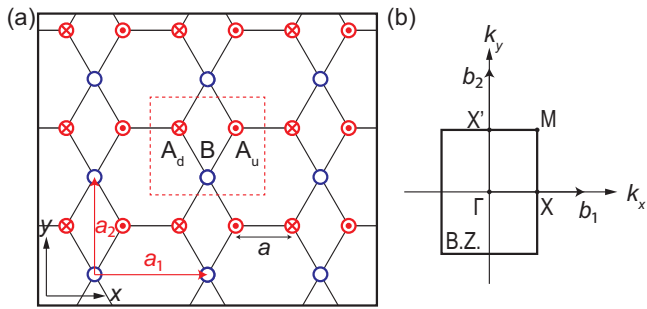}
\caption{\label{toymodel}(Color online)
Schematic figure of our toy model.
(a) Spring-mass model with localized spins, and (b) its first Brillouin zone.
The red circles with dots (${\rm A_u}$) and those with crosses (${\rm A_d}$) represent particles with spins perpendicular to $xy$ plane and effective negative charges, and the blue circles (B) represent particles with effective positive charges.
The primitive vectors are $\bm{a}_{1} = a(2,0), \bm{a}_2 = a(0,\sqrt{3})$ with lattice constant $a$.
The straight segments in (a) represent springs.
}
\end{figure}

In order to calculate the phonon angular momentum generated by an electric field, we introduce a toy model. 
Our toy model is a spring-mass model with localized spins, and this model is shown in Fig.~\ref{toymodel}.
In Fig.~\ref{toymodel}(a), the red circles with dots and those with crosses represent particles with effective negative charges $q_{\rm A} = -q_0/2$ and spins perpendicular to the $xy$ plane $\bm{S}=(0,0,\pm S)$, respectively, 
 and the blue circles represent particles with effective positive charges $q_{\rm B} = q_0$, where $q_0$ is a positive constant. 
This model has three particles in the unit cell. 
The motions of the particles are confined within the $xy$ plane. 
Thus, this model represents an antiferromagnetic insulator. 
Here, the primitive vectors are $\bm{a}_{1} = a(2,0), \bm{a}_2 = a(0,\sqrt{3})$ with lattice constant $a$, and the reciprocal vectors are $\bm{b}_{1} = (2\pi/a)(1/2,0)$, $\bm{b}_{2} = (2\pi/a)(0,1/\sqrt{3})$. 
The first Brillouin zone is shown in Fig.~\ref{toymodel}(b). 
We label the particles 
represented by the red circles with dots, those with crosses and blue circles  as ${\rm A_{u}, A_{d} }$, and ${\rm B}$, respectively. 
The spring constants of the ${\rm A_{u}}$-${\rm A_{d}}$ bonds and the \rm{A-B} bonds are $k_1$ and $k_2$, respectively. 
The potential energy of the springs is represented as
\begin{equation}
U = \frac{1}{2}\sum_{\braket{l\kappa,l^\prime \kappa^\prime}}k_{l\kappa,l^\prime\kappa^\prime}(|\bm{R}_{l\kappa}-\bm{R}_{l^\prime \kappa^\prime} + \bm{u}_{l\kappa}-\bm{u}_{l^\prime \kappa^\prime}|-l_{l\kappa,l^\prime\kappa^\prime})^2, 
\end{equation}
where $\bm{R}_{l\kappa}$ and $\bm{u}_{l\kappa}$ are the equilibrium position vectors and the displacement vector of the $\kappa$th particle in the $l$th unit cell, respectively, and $k_{l\kappa,l^\prime\kappa^\prime}$ and $l_{l\kappa,l^\prime\kappa^\prime}$ are the spring constants and the length of the springs between the $\kappa$th particle in the $l$th unit cell and the $\kappa^\prime$th particle in the $l^\prime$th unit cell, respectively. $\braket{l\kappa,l^\prime\kappa^\prime}$ represents a pair of the nearest neighbor particles. 
For simplicity, we assume that the Coulomb potential due to the effective charge is included in the spring constant.
We set the spring constants $k_1, k_2$ to be equal, $k_1=k_2=k$ and the length of the all springs $l_0$.
The Lagrangian for phonons is represented as
\begin{align}
L &= 
\sum_{l\kappa}\frac{1}{2}m_{\kappa}\bm{u}_{l\kappa}^2 
- \sum_{l\kappa}\sum_{l^\prime \kappa^\prime} \frac{1}{2} \Phi_{\alpha\beta}(l\kappa,l^\prime \kappa^\prime) u_{l\kappa,\alpha}u_{l^\prime \kappa^\prime,\beta} \notag \\
&\hspace{2.5cm} +  \sum_{l\kappa} g \bm{S}_{\kappa}\cdot (\bm{u}_{l\kappa}\times m_{\kappa}\dot{\bm{u}}_{l\kappa}), \\
&\Phi_{\alpha\beta}(l\kappa,l^\prime \kappa^\prime) = \frac{\partial^2 U}{\partial u_{l\kappa,\alpha} \partial u_{l^\prime \kappa^\prime, \beta}} \Biggr |_{u\to 0},
\end{align}
where $\Phi$ is the force constant matrix, $m_{{\rm A} ({\rm B})}$ is a mass of the particle ${\rm A (B)}$. 
To make the formula compact, we rewrite the displacement vectors within the unit cell into a six-dimensional vector represented as 
\begin{equation}
\bm{u}_{l\kappa} \to u_{l} = (\sqrt{m_{\rm A}}\bm{u}_{l {\rm A_{u}}},\sqrt{m_{\rm A}}\bm{u}_{l {\rm A_{d}}},\sqrt{m_{\rm B}}\bm{u}_{l {\rm B}})^{T}.
\end{equation}
The Hamiltonian for the phonons in this toy model is represented as 
\begin{align}
H 
&= \frac{1}{2}\sum_{l}\left[ (p_{l}-\beta_{z}u_l)^T(p_{l}-\beta_z u_l) + \sum_{l^\prime} u_l^T \Phi_{ll^\prime}u_l^\prime\right], \\
\beta_z
&= g
\begin{pmatrix}
S & 0 & 0 \\
0 & -S & 0 \\
0 & 0 & 0
\end{pmatrix}
\otimes
\begin{pmatrix}
0&-1\\
1&0
\end{pmatrix}, \\
\Phi_{ll^\prime}
&=
\begin{pmatrix}
\frac{\Phi(l{\rm A_{u}},l^\prime {\rm A_{u}})}{m_{\rm A}} & \frac{\Phi(l{\rm A_u},l^\prime {\rm A_d})}{m_{\rm A}} & \frac{\Phi(l{\rm A_{u}},l^\prime {\rm B})}{\sqrt{m_{\rm A} m_{\rm B}}} \\
\frac{\Phi(l{\rm A_{d}},l^\prime {\rm A_{u}})}{m_{\rm A}} & \frac{\Phi(l{\rm A_{d}},l^\prime {\rm A_{d}})}{m_{\rm A}} & \frac{\Phi(l{\rm A_d},l^\prime {\rm B})}{\sqrt{m_{\rm A} m_{\rm B}}} \\
\frac{\Phi(l{\rm B},l^\prime {\rm A_{u}})}{\sqrt{m_{\rm B}m_{\rm A}}} & \frac{\Phi(l{\rm B},l^\prime {\rm A_d})}{\sqrt{m_{\rm B} m_{\rm A}}} &\frac{\Phi(l{\rm B},l^\prime {\rm B})}{m_{\rm B}} 
\end{pmatrix}.
\end{align}

The symmetry of this toy model is characterized by the magnetic point group $mm^\prime m$, generated by $\sigma_{z},C_{2y},{\rm TR}\times I$, where TR is the time-reversal operation. 
We note that the prime means a joint operation of a spatial operation and the time-reversal operation TR, and we use the notation of Bilbao Crysallographic Server~\cite{Aroyo1, Aroyo:xo5013}.
In equilibrium, the phonon angular momentum for each phonon mode becomes zero at all wave vectors due to the ${\rm TR}\times I$ symmetry, and therefore the total phonon angular momentum is zero. 
We consider the response tensor $\beta$ of the phonon rotoelectric effect in Eq.~(\ref{pmee}). 
From symmetry analysis for $mm^\prime m$, nonzero elements of the response tensor $\beta_{\alpha \beta}$  are $\beta_{xz}$ and  $\beta_{zx}$. 
This form of the response tensor $\beta$ can also be seen from the following argument. 
For example, when the electric field is applied along the $x$-direction, the equilibrium positions of particles are slightly shifted along the $x$-direction due to the electrostatic force, and thus, the symmetry of this system is lowered to $2^\prime m^\prime m$, which necessarily leads to a nonzero angular momentum along the $z$-direction. 
On the other hand, when the electric field is applied along the $y$-direction, the equilibrium positions of particles are slightly shifted along the $y$-direction, and the symmetry of this system becomes $m2m$ which is a magnetic point group generated by $\sigma_x$ and $\sigma_z$. 
The phonon angular momentum along the $z$-direction cancels between $(k_x,k_y)$ and $(-k_x,k_y)$ due to the $C_{2y}$ symmetry. Therefore, when the electric field is applied along the $x$-direction, the phonon angular momentum is generated and it is proportional to the electric field.

\section{Numerical results and discussion}
We calculate the phonon angular momentum generated by the electric field. 
In this toy model, let $\tilde{\bm{d}}_{\kappa}$ denote the slight deviation of the 
equilibrium position of the $\kappa$-th particle in the unit cell along the external electric field $\bm{E}$. 
This lattice deformation comes from the electrostatic force onto the effective charges of the particles. 
In the deformed lattice structure, the new equilibrium positions become $\bm{R}_{l\kappa}^{\prime} = \bm{R}_{l\kappa} + \tilde{\bm{d}}_{\kappa}$. 
The potential energy of the spring is rewritten as
\begin{align}
U 
= \frac{1}{2}\sum_{\braket{l\kappa,l^\prime \kappa^\prime}}k(|\bm{R}_{l\kappa}^\prime-\bm{R}_{l^\prime \kappa^\prime}^\prime + \bm{u}_{l\kappa}-\bm{u}_{l^\prime \kappa^\prime}|-l_0)^2, 
\end{align}
To satisfy the balance of the force onto each particle at its equilibrium position, the relation between the deviation $\tilde{\bm{d}}=\tilde{\bm{d}}_{\rm B} - \tilde{\bm{d}}_{\rm A}$ and the external electric field $\bm{E}$ is represented as 
\begin{equation}
(E_x,E_y) = -\frac{k}{q_0}((4-3\eta)\tilde{d}_x,(4-\eta)\tilde{d}_y) \label{Evsd} ,
\end{equation}
with a parameter $\eta = l_0/a$. 

Here, we show the numerical result of our toy model. 
For simplicity, we set the parameters as $k=1$, $m_{\rm A} = m_{\rm B} = 1$, $g = 0.1$, $S = 1$, and $\eta = 0.8$. 
When the restoring force does not work $(\eta = 1)$, the phonon dispersion has zero-frequency flat band and this model is unstable. Therefore we set $\eta < 1$.

\begin{figure*}
\includegraphics[width=16cm]{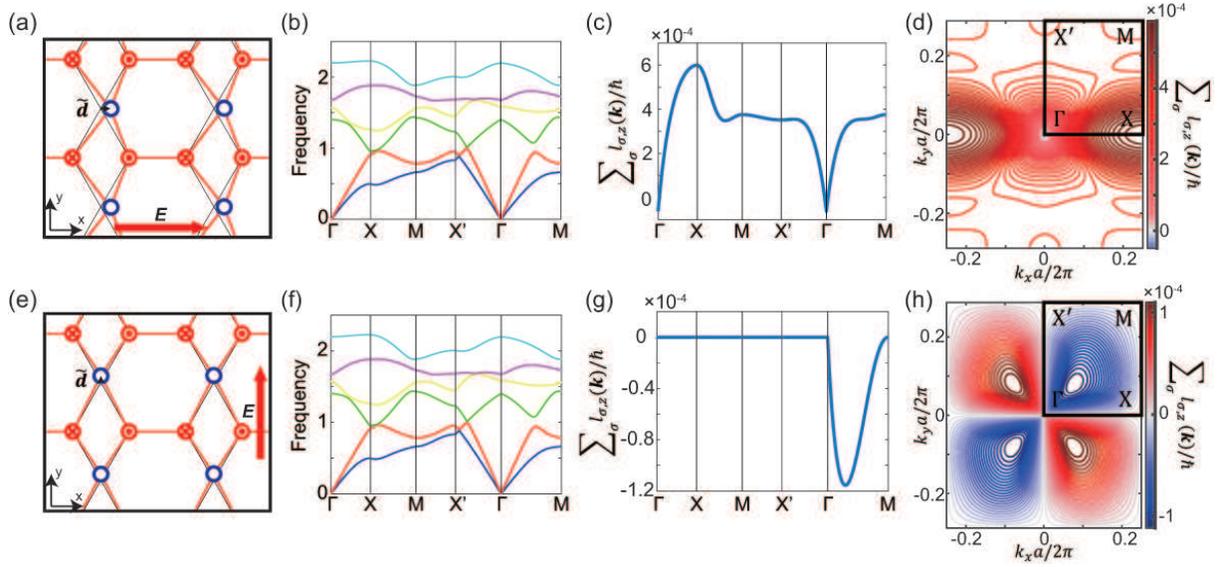}
\caption{\label{pdandpamwithE}(Color online)
Phonon dispersion and the phonon angular momentum with the electric field.
(a) and (e): Deformed lattice structure by the electric field along the $x$-direction and along the $y$-direction, respectively. 
(b) and (f): Phonon dispersion with the electric field $E_x$ and with the electric field $E_y$, respectively.
(c) and (g): Sum of the phonon angular momentum over the phonon modes with the electric field $E_x$ and with the electric field $E_y$ on the high-symmetry lines, respectively.
(d) and (h): Sum of the phonon angular momentum over the phonon modes with the electric field $E_x$ and with the electric field $E_y$, shown in the first Brillouin zone, respectively.
We set the parameters as $k = 1$, $m_{\rm A} = m_{\rm B} = 1$, $\eta = 0.8$, and $\tilde{\bm{d}}=(a/50,0)$.
}
\end{figure*}

First, we show the phonon dispersion and phonon angular momentum in the system with the electric field along the $x$-direction $E_x$. 
In Figs.~\ref{pdandpamwithE}(a) and (b), we show the deformed lattice structure by the electric field $E_x$ and its phonon dispersion, respectively. 
In Figs.~\ref{pdandpamwithE}(c) and (d), we show the sum of the phonon angular momentum along the $z$-direction of all the phonon modes with the electric field $E_x$ on the high-symmetry line and in the first Brillouin zone, respectively. 
The sum of the phonon angular momentum along the $z$-direction becomes nonzero. 
Due to ${\rm TR}\times C_{2x}$ and ${\rm TR}\times \sigma_{y}$ symmetries, the phonon angular momentum satisfies $l_{\sigma,z}(k_x,k_y) = l_{\sigma,z}(-k_x,k_y)$, and then the total phonon angular momentum along the $z$-direction can be nonzero even after the summation over the all the phonon modes.
At finite temperature, the total phonon angular momentum is given by the sum of the product between $l_{\sigma,z}(\bm{k})$ and the Bose distribution function, and it is also nonzero.

Next, we show the phonon dispersion and phonon angular momentum in the system with the electric field along the $y$-direction $E_y$. 
In Figs.~\ref{pdandpamwithE}(e) and (f), we show the deformed lattice structure by the electric field $E_y$ and its phonon dispersion, respectively. 
In Figs.~\ref{pdandpamwithE}(g) and (h), we show the sum of the phonon angular momentum along the $z$-direction of all the phonon modes with the electric field $E_y$ on the high-symmetry line and in the first Brillouin zone, respectively. 
The sum of the phonon angular momentum along the $z$-direction cancels between $(k_x,k_y)$ and $(-k_x,k_y)$ due to $\sigma_x$ and $C_{2y}$ symmetries. 
Therefore, the phonon angular momentum with the electric field $E_y$ have the nonzero value at any wave vector $\bm{k}$, and its sum over the wave vector in the first Brillouin zone vanishes.
We note that the dipersions with $\bm{E}\parallel \hat{x}$ (Figs.~\ref{pdandpamwithE}(b)) and $\bm{E}\parallel \hat{y}$ (Fig.~\ref{pdandpamwithE}(f)) look similar, but these two cases have different eigenstates, leading to different behaviors for angular momenta.

\begin{figure}
\includegraphics[width=6.5cm]{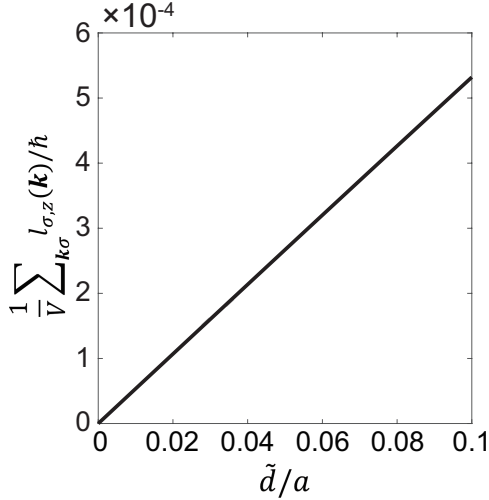}
\caption{\label{pamExlinear}
Dependence of the phonon angular momentum on the atomic deviation $\tilde{d}$ due to the electric field $E_x$.}
\end{figure}

Next, we show dependence of the phonon angular momentum under the electric field $E_x$ on the atomic deviation $\tilde{d}$ in Fig.~\ref{pamExlinear}. 
The sum of the phonon angular momentum of all the modes over the wave vector in the first Brillouin zone is proportional to the slight deviation $\tilde{d}$. 
Because, from Eq.~(\ref{Evsd}), the slight deviation $\tilde{d}$ is proportional to the electric field $\bm{E}$, the phonon angular momentum is proportional to the electric field.
Thus, this gives  microscopic mechanism for the phonon rotoelectric effect.
From this result, it follows that 
applying electric fields in opposite directions
reverses the handedness of elliptical motion of the ions.

Next, we discuss the phonon rotoelectric effect and the phonon thermal Edelstein effect in a three-dimensional system in the high-  and the low-temperature limit. 
From the discussion above, the response tensor $\beta_{\alpha\beta}$ for the phonon rotoelectric effect behaves similarly to the equilibrium phonon angular momentum in a system without  time-reversal symmetry in the high- and the low-temperature limit. 
Therefore, as is similar to the previous study
\cite{PAM01},
the phonon rotoelectric effect tensor $\beta_{\alpha\beta}$ vanishes in the high-temperature limit.
On the other hand, in the low-temperature limit, 
the zero-point motion becomes dominant for the phonon angular momentum.
As shown in the appendix, 
the phonon rotoelectric tensor $\beta_{\alpha\beta}$ 
in the low-temperature limit has a constant term due to zero-point motion and a term proportional to $T^5$ coming from the Bose distribution. 

On the other hand, 
as shown in the appendix, the phonon angular momentum by the phonon thermal Edelstein effect becomes constant in the high-temperature limit, and becomes proportional to $T^3$ in the low-temperature limit. 
These differences in temperature dependence between the phonon rotoelectric effect and the phonon thermal Edelstein effect 
come from the difference in their physical origins.
In the phonon thermal Edelstein effect, the deviation 
of the phonon distribution from equilibrium contributes to the total phonon angular momentum.
On the other hand, in the phonon rotoelectric effect, the change of the phonon angular momentum by the reduction of
the symmetry contributes.
Therefore, the temperature dependences of these effects are quite different.

We compare the two effects of generation of phonon angular momentum.
One is by an external magnetic field, as proposed in Ref.~\cite{PAM01}, and the other is by an external electric field, i.e. the phonon rotoelectric effect.
Both in these two effects, the spin-phonon interaction Eq.~(\ref{spi}) is needed.

First, we roughly estimate the response to the external magnetic field $\bm{B}$, the system acquires a spin angular momentum $V_{\rm cell}\chi B$, where $\chi$ is the susceptibilitiy and $V_{\rm cell}$ is the volume of unit cell.
This gives rise to a spin-phonon coupling of the size $\sim gV_{\rm cell}\chi B\hbar$.
Compared with the phonon bandwidth, i.e. the Debye frequency $\omega_{\rm D}$, its dimensionless size is $gV_{\rm cell}\chi B\hbar/(\hbar \omega_{\rm D}) = gV_{\rm cell}\chi B/\omega_{\rm D}$. 
Thus we evaluate the phonon angular momentum to be $(gV_{\rm cell}\chi B/\omega_{\rm D})\times \hbar $ per mode. 
Thus the total phonon angular momentum per unit volume is $J_{z}^{\rm ph} \sim gV_{\rm cell}\chi B\hbar/(\omega_{\rm D} V_{\rm cell}) = g\chi B\hbar/\omega_{\rm D}$.

Next, we roughly evaluate the response to the electric field $\bm{E}$.
The electric field $E$ induces a polarization $(\varepsilon-\varepsilon_0) E$, where $\varepsilon$ is the dielectric constant and
$\varepsilon_0$ is the vacuum permittivity.
Then the displacement is of the order of $(\varepsilon-\varepsilon_0) E V_{\rm cell}/e$, where $-e$ is the electronic charge.
This modifies the dynamical matrix of the ratio $(\varepsilon-\varepsilon_0) E V_{\rm cell}/(ea)$, where $a$ is the lattice constant.
This perturbation to the dynamical matrix, together with the spin-phonon interaction whose diminsionless size is $g\hbar/\omega_{\rm D}$, the phonon angular momentum is $((\varepsilon-\varepsilon_0)E V_{\rm cell}/(ea))\times(g\hbar/\omega_{\rm D})\times \hbar = (\varepsilon-\varepsilon_0) E V_{\rm cell} g\hbar^2/(e a \omega_{\rm D})$.
Thus the total phonon angular momentum per unit volume is $J_z^{\rm ph} \sim ((\varepsilon-\varepsilon_0) E V_{\rm cell} g \hbar^2/(ea\omega_{\rm D}))/V_{\rm cell} = (\varepsilon-\varepsilon_0)E g\hbar^2/(ea\omega_{\rm D})$.
Using this result,
we estimate the magnitude of the angular momentum induced by an electric field
for a real material. Thus far, we do not have quantitative estimate for the spin-phonon coupling constant $g$ for a real material, and  
here we set it to be 1 cm$^{-1}/\hbar^2$, as adopted in Ref.~\onlinecite{Sheng}. 
We consider Cr$_2$O$_3$ as an example for a multiferroic material, and its 
phonon frequency is typically $\omega_D \sim 10^{1}\mathrm{THz}\sim10^{-2}$eV \cite{WangSurfSci}.
Therefore, $g\hbar/\omega_D\sim 1\mathrm{cm}^{-1}/10^{-2}\mathrm{eV}\sim10^{-3}$. 
On the other hand, the displacement of atoms is around 10$^{-15}$m for an
electric field $\sim10$ V/mm, by using  the dielectric constant $\varepsilon\sim 11
\varepsilon_0$ \cite{Cr2O3-dielec} and the lattice constant $\sim 5$\AA.
Thus, the induced angular momentum per each mode is of the order $10^{-15}\mathrm{m}/(5\mbox{\AA})\times10^{-3}\times \hbar
\sim 10^{-8}\hbar$.
At room temperature $T=300$K, the phonon population is of the order $k_BT/\hbar\omega_D\sim 1$, and
therefore the induced angular momentum per unit cell at room temperature is
also of the order $\sim 10^{-8}\hbar$.
Note that it is only an order estimate, and the effect of the external field depends on phonon modes.

\section{Conclusion}
We have theoretically predicted generation of the phonon angular momentum 
by an electric field in systems without time-reversal and inversion symmetries. 
When time-reversal and inversion symmetries are broken and their product is conserved, the phonon angular momentum vanishes in equilibrium, and it is generated by an electric field. 
We note that in an ionic crystal, the phonon
rotoelectric effect automatically induces 
a phonon magnetoelectric effect.
This mechanism is analogous to the magnetoelectric effect in multiferroic materials as is expected from symmetry. 
We calculate the generated phonon angular momentum using a toy model.

Since the phonon rotoelectric effect occurs under the same conditions of symmetry as in the magnetoelectric effect, we expect that  the phonon angular momentum is generated by an electric field in multiferroic materials with the spin-phonon interaction. 
However, in general, the spin-phonon interaction is weak, and thus the phonon angular momentum by this effect may be small.
Similar to our previous work~\cite{Masato02}, when the sample is suspended by a string so that it can freely rotate, the angular momentum is transferred into phonons, electrons, and the rigid-body rotation of the crystal due to the conservation of the angular momentum.
Only in ionic crystals where the nuclei have nonzero effective charge, the phonon angular momentum by this effect also leads to magnetization.

We also expect its inverse effect, a creation of electric polarization by exciting or injecting a phonon angular momentum in the sample from a symmetry viewpoint.
Nonetheless it should occur off the equilibrium, and its microscopic mechanism should be quite different from the phonon rotoelectric effect.

\begin{acknowledgments}
This work was partly supported by a MEXT KAKENHI Grant Number JP
26100006, by JST CREST Grant Number JPMJCR14F1, and also by JSPS KAKENHI Grant Number JP17J10342.
\end{acknowledgments}

\appendix*
\section{Phonon angular momentum in the high- and the low-temperature limits}
In this section, we derive behaviors of the phonon thermal Edelstein effect and the phonon rotoelectric effect in a three-dimensional system in the high- and the low-temperature limits. 
We first consider the phonon thermal Edelstein effect. 
At high temperature $\hbar \omega_{\sigma}(\bm{k}) \ll k_{\rm B}T$, we can expand the Bose distribution function as 
\begin{align}
f_0(x) = \frac{1}{e^x-1} = \frac{1}{x} - \frac{1}{2} + \frac{1}{12}x + O(x^2),
\end{align}
with $x = \frac{\hbar \omega}{k_{\rm B}T} (\ll 1)$. Then, the temperature derivative  of the Bose distribution function is represented as 
\begin{equation}
\frac{\partial f_0(x)}{\partial T} 
\simeq \frac{k_{\rm B}}{\hbar \omega} - \frac{\hbar \omega}{12 k_{\rm B}T^2}.
\end{equation}
The response tensor $\alpha$ for the phonon thermal Edelstein effect in the high-temperature limit is represented as
\begin{equation}
\alpha_{\alpha\beta} = -\frac{\tau}{V} \sum_{\bm{k},\sigma} l_{\sigma,\alpha}v_{\sigma,\beta}(\bm{k})\frac{k_{\rm B}}{\hbar\omega_{\sigma}(\bm{k})},
\end{equation}
where $\bm{v}_{\sigma}(\bm{k}) = \frac{\partial \omega_{\sigma}(\bm{k})}{\partial \bm{k}}$ is the group velocity of a phonon mode.
This shows that the response tensor becomes constant in the high-temperature limit.

At low temperature $\hbar \omega_{\sigma}(\bm{k}) \gg k_{\rm B}T$, we consider only the acoustic modes with a long wave length, because populations in the phonon modes except for the acoustic modes are negligibly small. In the long wave length limit, the frequencies of acoustic modes are represented as
\begin{equation}
\omega_{\sigma}(\bm{k}) = v_{\sigma} \sqrt{k_x^2 +k_y^2 +k_z^2} = v_{\sigma}k,
\end{equation}
where $\sigma = 1,2,3$ represents a band index for the acoustic modes. For simplicity, we assume that the group velocity is isotropic and we change the wave vector $\bm{k}$ from the Cartesian coordinate $(k_x,k_y,k_z)$ to the polar coordinate $(k,\theta,\phi)$. 
We can expand the phonon angular momentum of each mode in the Taylor series as
\begin{equation}
l_{\sigma,\alpha}(\bm{k}) = \gamma_{\sigma, \alpha \beta}k_{\beta} + O(k^2).
\end{equation}
Since we are focusing on systems without inversion symmetry, linear order terms are allowed here.
The coefficient tensor $\gamma_{\alpha\beta}$ is an axial tensor because the phonon angular momentum of each mode is an axial vector and the wave vector is a polar vector. In the system of an infinite size, the wave vectors are continuous, so that one can replace summation by integration,
\begin{equation}
\frac{1}{V}\sum_{\bm{k}} \to 
\frac{1}{(2\pi)^3}\int_0^{2\pi}d\phi \int_0^\pi d\theta \int_0^{k^{c}}dk k^2 \sin\theta,
\end{equation}
where $k^{c}$ is a cutoff wavenumber introduced for the short wavelength. Then, the response tensor $\alpha$ can be rewritten as
\begin{align}
\alpha_{\alpha\beta} 
&= -\frac{\tau}{V}\sum_{\sigma=1,2,3}\sum_{\bm{k}}l_{\sigma,\alpha}(\bm{k})v_{\sigma,\beta}(\bm{k})\frac{\partial f_0(\omega_{\sigma}(\bm{k}))}{\partial T} \notag \\
&= \sum_{\sigma=1,2,3} \frac{-\tau v_{\sigma} \gamma_{\sigma,\alpha\beta}}{6\pi^2} \int_{0}^{k^{c}} dk k^3 \frac{\partial f_0(\omega_{\sigma}(k))}{\partial T}.
\end{align}
Here, we use the following relation
\begin{equation}
\frac{\partial f_0(\omega_\sigma(k))}{\partial T} = -\frac{\omega_\sigma}{T}\frac{\partial f_0(\omega_\sigma(k))}{\partial \omega_{\sigma}},
\end{equation}
and we introduce the cutoff frequency $\omega_{\sigma}^{c}=v_{\sigma}k^{c}$. Then, the response tensor is represented as
\begin{align}
\alpha_{\alpha\beta} 
&= \sum_{\sigma=1,2,3} \frac{\tau v_\sigma \gamma_{\sigma,\alpha\beta}}{6\pi^2 T} \int_0^{k^c} dk k^3 \omega_{\sigma} \frac{\partial f(\omega_{\sigma})}{\partial \omega_{\sigma}} \notag \\
&= \sum_{\sigma=1,2,3} \frac{\tau \gamma_{\sigma,\alpha\beta}}{6\pi^2 v_{\sigma}^3 T} \int_{0}^{\omega_{\sigma}^{c}} d\omega_{\sigma} \omega_{\sigma}^{4} \frac{\partial f_0(\omega_{\sigma})}{\partial \omega_{\sigma}} \notag \\
&= \sum_{\sigma = 1,2,3} \frac{\tau \gamma_{\sigma,\alpha\beta} k_{\rm B}^4 T^3}{6\pi^2v_{\sigma}^3 \hbar^4} 
\int_0^{x_{\sigma}^{c}} dx_{\sigma} x_{\sigma}^4 \frac{\partial f_0(x_{\sigma})}{\partial x_{\sigma}}. \label{pteelow1}
\end{align}
Because $x_{\sigma}^{c} = \frac{\hbar \omega_{\sigma}^{c}}{k_{\rm B}T} \to \infty$ in the low-temperature limit, the integral part in Eq.~(\ref{pteelow1}) can be calculated as 
\begin{align}
\int_0^{x_{\sigma}^c} dx_{\sigma} x_{\sigma}^4 \frac{\partial f_0(x_{\sigma})}{\partial x_{\sigma}}
&= -4 \int_0^{\infty} dx_{\sigma} \frac{x_{\sigma}^3}{e^{x_{\sigma}}-1} \notag \\
&= -24\zeta(4) \notag \\
&= -\frac{4\pi^4}{15}.
\end{align}
Therefore, the response tensor $\alpha$ is represented as
\begin{equation}
\alpha_{\alpha\beta} = -\sum_{\sigma=1,2,3} \frac{2\pi^2 \tau \gamma_{\sigma,\alpha\beta} k_{\rm B}^4}{45 v_{\sigma}^3 \hbar^4}T^3
\end{equation}
This shows that the response tensor is proportional to $T^3$ in the low-temperature limit.

On the other hand, we consider the behavior of the phonon rotoelectric effect. 
At high temperature, $\beta_{\alpha\beta}$ vanishes as $T^{-1}$, because 
the phonon angular momentum for the phonon rotoelectric effect at high temperature is represented as 
\begin{align}
J_{\alpha}^{\rm ph} = \frac{1}{V} \sum_{\bm{k},\sigma} l_{\sigma,\alpha}(\bm{k}) 
\left[
\frac{k_{\rm B}T}{\hbar\omega_{\sigma}(\bm{k})} + \frac{\hbar\omega_{\sigma}(\bm{k})}{12k_{\rm B}T}
\right],
\end{align}
and the first term vanishes due to $\sum_{\bm{k},\sigma}(l_{\sigma,\alpha}(\bm{k})/\omega_{\sigma}(\bm{k}))=0$
\cite{PAM01}.
On the other hand, at low temperature under an electric field, the phonon angular momentum which is the product between $l_{\sigma,\alpha}(\bm{k})$ and the Bose distribution function is represented as 
\begin{align}
\frac{1}{V}\sum_{\sigma = 1,2,3} \sum_{\bm{k}} l_{\sigma,\alpha}(\bm{k}) f_0(\omega_{\sigma}(\bm{k}))
= \sum_{\sigma=1,2,3} \frac{4\zeta(5)\delta_{\sigma,\alpha}}{\pi^2} \left(\frac{k_{\rm B}T}{\hbar v_{\sigma}} \right)^5,
\end{align}
where the phonon angular momentum for each phonon modes is represented as $l_{\sigma,\alpha} \sim \gamma_{\sigma,\alpha\beta}k_{\beta} + \delta_{\sigma,\alpha\beta\gamma}k_\beta k_{\gamma}$, and the coefficient tensor $\delta_{\alpha\beta\gamma}$ is an axial tensor and $\delta_{\sigma,\alpha} = \delta_{\sigma,\alpha\beta\beta}$.
Therefore, the phonon angular momentum due to the phonon rotoelectric effect in the low-temperature limit is represented as
\begin{align}
J_{\alpha}^{\rm ph} = \frac{1}{V}\sum_{\bm{k},\sigma} \frac{l_{\sigma,\alpha}(\bm{k})}{2} + \sum_{\sigma=1,2,3} \frac{4\zeta(5)\delta_{\sigma,\alpha}}{\pi^2} \left(\frac{k_{\rm B}T}{\hbar v_\sigma} \right)^5.
\end{align}
This shows that the response tensor $\beta_{\alpha\beta}$ for the phonon rotoelectric effect is also proportional to $T^5$ in the low-temperature limit.



%










\end{document}